\documentclass[prb,preprint]{revtex4}

\pdfoutput=1

\usepackage{graphicx}
\usepackage{gensymb}

\begin{document}


\title{Unidirectionally shifted shadow arcs in  $\mathbf{La_{2-x}Sr_xCuO_4}$ explained}



\medskip 

\date{July 18, 2017} \bigskip

\author{Manfred Bucher \\}
\affiliation{\text{\textnormal{Physics Department, California State University,}} \textnormal{Fresno,}
\textnormal{Fresno, California 93740-8031} \\}

\begin{abstract}
The observed faint shadow arcs, shifted unidirectionally from primary Fermi arcs, are caused by lattice distortion from edge dislocations introduced by cleaving the samples.

\end{abstract}

\maketitle

\pagebreak

Angle-resolved photoemission spectrocopy (ARPES) of nearly optimally doped $La_{2-x}Sr_xCuO_4$ ($x = 0.15, 0.17$) has revealed faint shadow arcs that are duplicates of the primary Fermi arcs shifted by $\mathbf{q_a} = (\pi, 0)$.\cite{1} No shadow arcs were found, however, shifted by $\mathbf{q_b} = (0, \pi)$.\cite{1} 
In the authors' opinion the observation could point to a new instability in the system that breaks rotation symmetry.\cite{1}
Both samples are in the doping range of orthorhombic crystal structure ($x < 0.21$).\cite{2} It is therefore surprising that shadow arcs appear only shifted by $\mathbf{q_a}$ but not also by $\mathbf{q_b}$. The purpose of this letter is to point out a \emph{unidirectional} influence on the crystals: the \emph{cleaving} of the samples. This raises the possibility that the unidirectional shift of the shadow arcs may be caused by sample preparation.

ARPES is a surface-sensitive technique. A freshly cleaved, plane surface is important. The samples under consideration were cleaved \textit{in situ} with an on-board sample cleaver,\cite{3} presumably in the $a$-direction (forward motion of the cleaver blade across the $ab$ cleavage plane in the crystal's $a$-direction).  The edge thickness of the cleaver blade ($D > 1 \mu m$) is at least three orders of magnitude larger than the spacing of the atomic layers of the $La_{2-x}Sr_xCuO_4$ crystal perpendicular to the $c$-direction ($\Delta c < 1 nm$). The squeeze of the cleaver blade on the sample, held in place by an anvil, introduces plastic deformations on the atomic scale, known as dislocations.\cite{4,5,6} Under the present condition these would be edge dislocations. One such edge dislocation in the $CuO_2$ plane is qualitatively shown in Fig. 1, characterized by its glide line in the $a$-direction and symmetry line in the $b$-direction (horizontal and vertical line, respectively). In order to understand the essential geometry of an edge dislocation, imagine that below the glide line ($b < 0$) two (because of charge neutrality) ion half-rows were removed from the virgin $CuO_2$ plane. Upon relaxation, ion positions below the glide line   and near the symmetry line are more diluted than in a virgin $CuO_2$ plane. Conversely, they are then denser above the glide line ($b > 0$). For the symmetry configuration shown in Fig. 1 alternate $Cu^{2+}$ and $O^{2-}$ ions reside on the symmetry line for $b > 0$ but only $O^{2-}$ ions for $b < 0$. The opposite holds for the other symmetry configuration (not shown). Far from the dislocation center (intersection of glide line and symmetry line) the ion displacements caused by the dislocation become negligible so that ion positions resemble again the virgin situation. Near the symmetry line the ion half-rows above and below the glide line are \emph{staggered} with a displacement of about half a lattice constant, $\Delta a \approx a_0/2$. The staggering and half-row displacement recede with increasing distance from the symmetry line such that half-rows join smoothly in the far region. There are many 

\pagebreak

\includegraphics[width=6in]{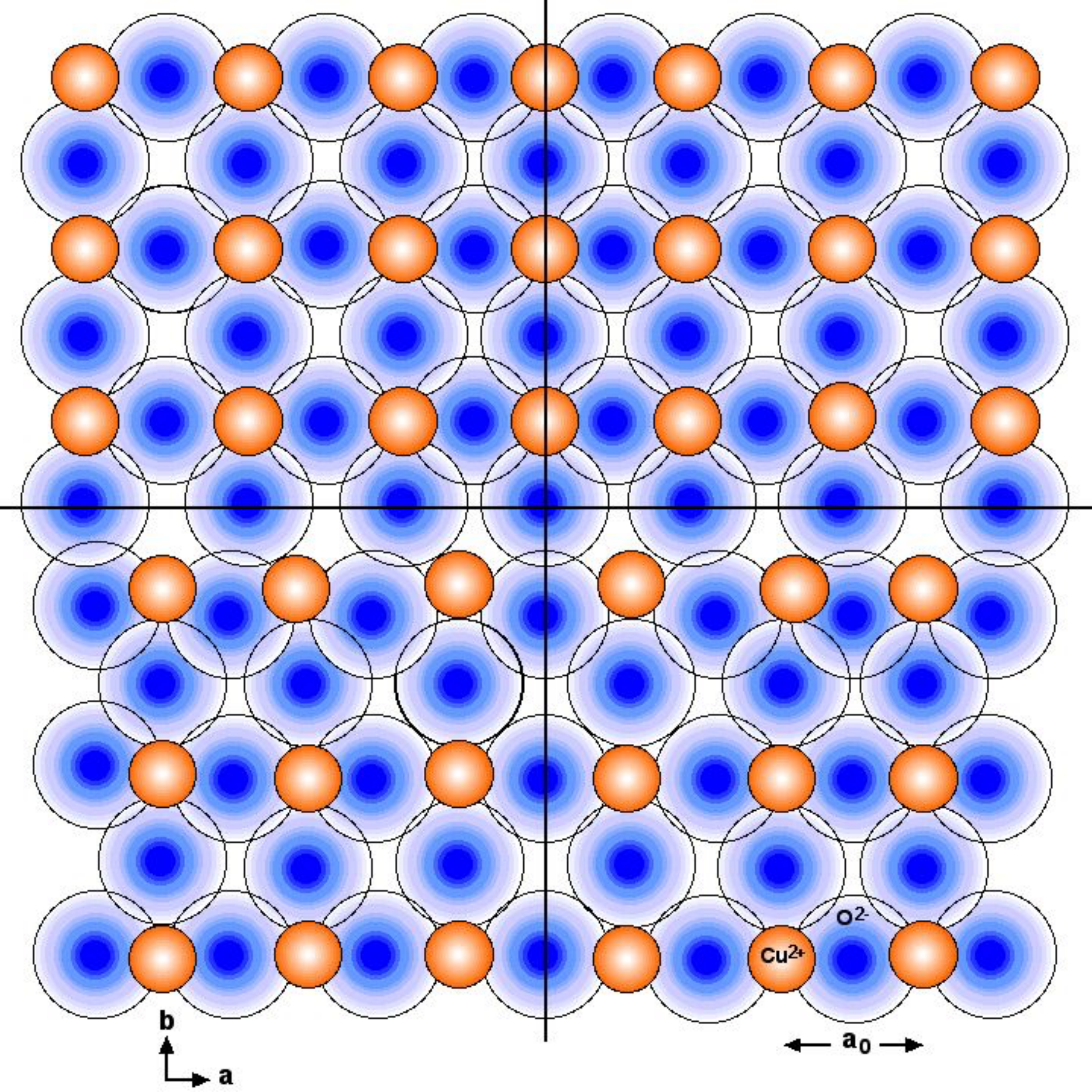}

\noindent FIG. 1. Edge dislocation (qualitatively) in the $CuO_2$ plane with (horizontal) glide line and (vertical) symmetry line. Ion positions are denser and, respectively, more diluted above and below the glide line. Near the symmetry line ion half-rows above and below the glide line are \emph{staggered} by about half a lattice constant, $\Delta a \approx a_0/2$. Far from the symmetry line (here, off-scale) ion half-rows join smoothly.
\pagebreak 

\noindent similarities of edge dislocations in the $CuO_2$ plane of a cuprate superconductor with the more studied case of edge dislocations in binary ionic crystals, such as NaCl.\cite{7}

Edge dislocations with glide lines solely in the $a$-direction are introduced by $a$-directional squeeze of the cleaver blade on the $La_{2-x}Sr_xCuO_4$  sample. The explosion (``crack'') of the cleaving occurs through the release of potential energy from piled-up dislocations. Thereafter some dislocations remain at, and near the cleaved surface. It is conjectured that lattice distortion from edge dislocations with $a$-directional glide lines, introduced by $a$-directional cleaving, give rise to the observed $\mathbf{q_a}$-shifted weak shadow arcs. 

If this conjecture is valid, it would have several consequences: 

(A) A sample, cleaved in the $a$-direction and then rotated about the $c$-axis by 90\degree \, in the ARPES spectrometer, would also show the shadow arcs shifted in the $\mathbf{q_a}$-direction (relative to the crystal) because the edge dislocations introduced by the cleaving will rotate with the sample. A $\mathbf{q_a}$-shift in the rotated sample has actually been observed.\cite{1}

(B) A sample cleaved in the $b$-direction would show weak shadow arcs that are exclusively $\mathbf{q_b}$-shifted.

(C) As $\mathbf{q_a}$-shifted shadow arcs are observed only near optimal doping (here, $x = 0.15, 0.17$), their doping dependence needs to be addressed. It can be assumed that plastic deformability of $La_{2-x}Sr_xCuO_4$ samples depends on the level of $Sr$ doping into the $LaO$ layers which corresponds to hole doping in
 the $CuO_2$ planes, $x = p$. If the deformability---and thereby the density of introduced edge dislocations by cleaving---increases with doping, this would qualitatively explain the onset of $\mathbf{q_a}$-shifted shadow arcs, say from \emph{very} weak at $x = 0.15$ to weak at $x = 0.17$. However, the \emph{absence}, instead of brighter shadow arcs at higher doping, confirmed for $x \ge 0.22$,\cite{1} must then be explained. No full explanation can be offered here, but a crucial change of circumstances should be pointed out. This is a quantum critical point at $x_0^* = 0.182$, marking the boundary of the pseudogap phase with the strange-metal phase at $T = 0$.\cite{8} Accompanying the quantum critical point in $La_{2-x}Sr_xCuO_4$ is a Lifshitz transition of the Fermi surface from hole-like to electron-like, as observed by ARPES.\cite{9} The (strange) metallicity, with its increased mobility of doped holes in the $CuO_2$ planes, may reduce the density of edge dislocation by lattice healing through reconstructive ion-row alignment.
 
(D) The ratio of spectral weight (``brightness'') of shadow arcs to primary Fermi arcs would correspond to the area ratio of dislocation-distorted regions in the cleaved surface to  
 
 \pagebreak

\includegraphics[width=6in]{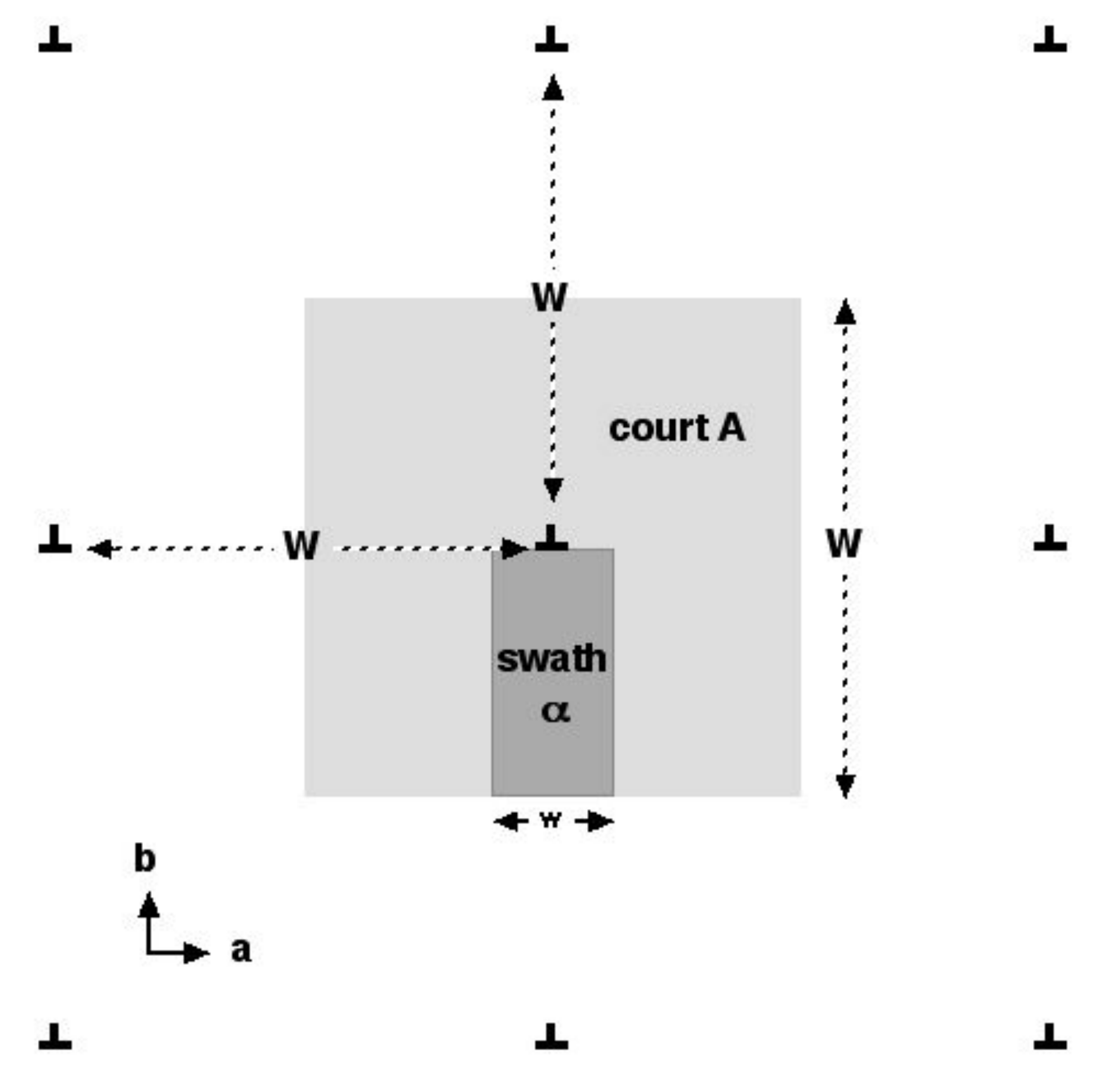}

\noindent FIG. 2. Division of an atomic $ab$ plane into square ``courts'' of edge length $W$ and area $A$ about centers $\bot$ on a square superlattice of lattice constant $W$. Each court contains a ``swath'' of area $\alpha$. 
The $\bot$ symbols show the \emph{average} position of dislocation centers.
As a \textit{toy model} of an edge dislocation with glide line in the $a$-direction,
ion rows are at ideal positions in the court, but shifted in the swath rigidly in the $a$-direction by half a lattice constant of the atomic configuration, $\Delta a = a_0/2$. In short, ion half-rows are ``staggered'' in the swath by $a_0/2$.

\pagebreak 

\noindent undistorted regions. The absence of shadow arcs in a sample that is not deformed (by being not cleaved) is, of course, not observable with the surface-reliant ARPES method. However, deliberate \emph{increase} of introduced edge dislocations by \emph{more} $a$-directional deformation of the sample (but below the cleaving threshold), before or after cleaving, should then give rise to \emph{brighter} shadow arcs.
 
(E) Conversely, healing the cleavage surface through thermal annealing or photon irradiation would \emph{dim} the brightness of the shadow arcs. If metallicity reduces the density of dislocations---a possibility raised in (C)---flooding the sample with free electrons by irradiation with photons of energy beyond the bandgap may achieve the same result.

(F) Plastic deformation of a sample in \emph{both} the $a$-direction and $b$-direction, before or after cleaving, should give rise to $\mathbf{q_a}$-shifted \emph{and} $\mathbf{q_b}$-shifted shadow arcs.

(G) The observed $\mathbf{q_a} =(\pi,0)$ shift of the shadow arcs should be explainable by the geometry of edge dislocations. For a derivation we make two approximations:

\noindent (i) An \emph{average} distribution of dislocations such that their centers ($\bot$) form a square superlattice in the $ab$ cleavage plane with edge length $W$ (see Fig. 2).

\noindent (ii) A \textit{toy model} of an edge dislocation, simplified such that all atoms (ions) reside at ideal lattice positions within a square \emph{court} of area $A$ about the center of each dislocation ($\bot$) except a \emph{swath}, of width $w$ and area $\alpha = w W/2$, below the glide line (see Fig. 2) where ion half-rows are rigidly displaced (``staggered'') in the $a$-direction by half a lattice constant, $\Delta a = a_0/2$.
This means that the toy model ignores compression and dilution above and below the glide line of a real edge dislocation. Instead it focuses exclusively on the $a_0/2$ staggering of ion half-rows, assumed only within the swath and ending abruptly at the swath-court interface (in contrast to a gradual transition in real edge dislocations).

For the present purpose it suffices to consider in the toy model only one ion species in the $CuO_2$ plane, say $Cu^{2+}$ ions at positions $\mathbf{R} = (ma_0, nb_0)$ \, ($m = ..., -2, -1, 0, 1, 2, ... \, ; \, n = ..., -\frac{3}{2}, -\frac{1}{2}, \frac{1}{2}, \frac{3}{2}, ...$) in the court, and at \emph{staggered} positions $\mathbf{R} = (ma_0 + a_0/2, nb_0)$ for $n < 0$ in the swath.
By Bloch's theorem the wavefunction of a crystal electron of momentum $\mathbf{p} = \hbar \mathbf{k}$ is a plane wave of wavevector $\mathbf{k}$, modulated by periodic atomic wavefunctions $u_\mathbf{k}(\mathbf{r})$,
\begin{equation}
\Psi_\mathbf{k}(\mathbf{r}) = u_\mathbf{k}(\mathbf{r}) \, exp(i \mathbf{k} \cdot \mathbf{r}) \, .
\end{equation}
The wavefunction in a large court, $W \gg a_0$, differs negligibly from that of an ideal crystal, $\Psi_\mathbf{k}(\mathbf{r})$.
For a large enough swath area the wavefunction  $\psi_\mathbf{k}(\mathbf{r})$ in its interior---sufficiently away from the discontinuity of the potential at the court-swath interface---will be proportional to the the wavefunction in the court, but displaced,
\begin{equation}
\psi_\mathbf{k}(\mathbf{r}) \propto \Psi_\mathbf{k}(\mathbf{r + \Delta r}) \, ,
\,\,\, \mathbf{\Delta r} = (\frac{a_0}{2}, 0) \,. 
\end{equation}
Leaving out the amplitude and factors common to $\psi_\mathbf{k}$ and $\Psi_\mathbf{k}$---such as $u_\mathbf{k}(\mathbf{r})$ and $exp(i k_b b)$---except those relevant to the staggered $\Delta a$ displacement, the swath wavefunction is reduced to
\begin{equation}
\psi_\mathbf{k}(\mathbf{r}) \propto exp(i k_a \frac{a_0}{2}) \, exp(ik_a a) \, . 
\end{equation}
Here the first exponential term is a phase factor, holding for \emph{all} positions $a$, due to the displacement of ion half-rows in the swath by half a lattice constant, $\Delta a = a_0/2$.

To proceed further we make the \textit{ansatz} that the wavefunction in the swath can equivalently be expressed through the Bloch wave from non-displaced ion positions---as in the court---but with a shift of \emph{wavevector},
\begin{equation}
\psi_\mathbf{k}(\mathbf{r}) \propto \Psi_\mathbf{k + \Delta k}(\mathbf{r}) \, ,
\,\,\, \mathbf{\Delta k} = (\Delta k_a, 0) \,. 
\end{equation}
Ignoring again common factors but those relevant to the $\Delta k_a$ shift, this reduces the swath wavefunction to
\begin{equation}
\psi_\mathbf{k}(\mathbf{r}) \propto exp(i \Delta k_a a) \, exp(ik_a a) \, . 
\end{equation}
Now the first exponential term is a phase factor that holds for all wavenumbers $k_a$.

The value of the wavenumber shift $\Delta k_a$ is determined by the special case $(\hat{k}_a, \hat{a}) = (k_{a0}, a_0)$ that satisfies both the conditions (3) and (5). It involves the lattice constant of $k$-space, $k_{a0} \equiv 2\pi/a_0$, and the 
lattice constant of the ideal ion configuration, $a_0$,
\begin{equation}
exp(i\frac{2\pi}{a_0} \frac{a_0}{2}) \, exp(i\frac{2\pi}{a_0} a_0) =
e^{i\pi} \, e^0 \equiv exp(i \Delta k_a a_0) \, exp(i \frac{2\pi}{a_0} a_0) =
e^{i \Delta k_a a_0} \, e^0 \, .
\end{equation}
Equality of the exponent in the second and forth (last) term yields
\begin{equation}
\Delta k_a = \frac{\pi}{a_0} \, ,
\end{equation}
or $q_a = \pi$ in reciprocal lattice units (rlu).

The plurality of the wavefunctions $\psi_{\mathbf{k}}(\mathbf{r})$ in the swath thus gives rise to quantum states (Bloch waves) displaced by half a lattice constant in real (ion configuration) space, $\Delta a = a_0/2$, and displaced in $k$-space by half a lattice constant of the reciprocal (wavenumber) lattice, $\Delta k_a = k_{a0}/2 = \pi/a_0$. 
The second scenario includes the quantum states of highest energy that comprise the shadow Fermi arcs, being shifted from the primary Fermi arcs by $\mathbf{q_a} = (\pi, 0)$, as observed.

(H) With the dislocation toy model, the brightness ratio $\beta/B$ between shadow arcs and primary Fermi arcs, qualitatively addressed in (D) and (E), can quantitatively be related to the court-swath geometry in terms of swath area $\alpha = wW/2$ and court area $A = W^2 - \alpha$ (see Fig. 2). This gives the relative brightness of shadow arc and primary arc,
 \begin{equation}
\frac{\beta}{B} = \frac{\alpha}{A} = \frac{wW/2}{W(W-w/2)} = \frac{w}{2W-w} \approx \frac{w}{2W} \, ,
\end{equation}
in terms of swath width $w$ and court width $W$.

(I) More generally, the observation of shadow patterns in $k$-space, by ARPES or other methods, shifted from primary patterns by half a lattice constant of reciprocal space, may serve as an indirect probe of edge dislocations in a sample.

\bigskip \bigskip 

\centerline{ \textbf{ACKNOWLEDGMENT}}

\noindent I thank Duane Siemens for valuable discussions and advice.


\begin{thebibliography}{9}

\bibitem{1}  E. Razzoli, C. E. Matt, Y. Sassa, M. M\r{a}nsson, O. Tjernberg, G. Drachuck, N. Monomo, M. Oda, T. Kurosawa, Y. Huang, N. C. Plumb, M. Radovic, A. Keren, L. Patthey, J. Mesot, and M. Shi, Phys. Rev. B \textbf{95}, 224504 (2017).

\bibitem{2} P. G. Radaelli, D. G. Hinks, A. W. Mitchell, B. A. Hunter, J. L. Wagner, B. Dabrowski, K. G. Vandervoort, H. K. Viswanathan, and J. D. Jorgensen, Phys. Rev. B \textbf{49}, 4163 (1994).

\bibitem{3} M. M\r{a}nsson, T. Claesson, U. O. Karlsson, O. Tjernberg, S. Pailh\'{e}s, J. Chang, J. Mesot, M. Shi, L. Patthey, N. Momono, M. Oda, and M. Ido, Rev. Sci. Instrum. \textbf{78}, 076103 (2007).

\bibitem{4} A. H. Cottrell, \textit{Dislocations and Plastic Flow in Crystals}, (Oxford UP, London, 1953).

\bibitem{5} A. Seeger, Handb. Phys., Vol. VII:2 (Springer, Berlin, 1958).

\bibitem{6} J. Friedel, \textit{Dislocations}, (Pergamon Press, Oxford, 1964).
 
\bibitem{7} M. Bucher, Phys. Status Solidi B \textbf{114}, 383 (1982).

\bibitem{8} M. Bucher, ``Fermi arc, pseudogap and strange-metal phase in hole-doped lanthanum cuprates'', arXiv:1704.05374v1

\bibitem{9}  E. Razzoli, Y. Sassa, G. Drachuck, M. M\r{a}nsson, A. Keren,
M. Shay, M. H. Berntsen, O. Tjernberg, M. Radovic, J. Chang, Pailh\'{e}s,  N. Monomo, M. Oda, M. Ido, O. J. Lipscombe, S. M. Hayden, L. Patthey, J. Mesot, and M. Shi, New J. Phys. \textbf{12}, 125003 (2010).

\end{thebibliography}
\end{document}